\documentclass[%
 reprint,
superscriptaddress,
showpacs, 
showkeys,
 amsmath,amssymb,
 aps,
]{revtex4-1}

\newcommand{\neb}{$\bar{\nu}_{e} \; $}
\newcommand{\nebns}{$\bar{\nu}_{e}$}

\newcommand{\ang}{$\theta _{13} $}
\newcommand{\sang}{$\rm{sin^2 2} \theta_{13} $}
\newcommand{\sangsp}{$\rm{sin^2 2} \theta_{13} \; $}

\newcommand{\Ep}{E$_{\rm prompt}$}
\newcommand{\Ed}{E$_{\rm delay}$}
\newcommand{\Delrel}{$\Delta t_{e^+n}$\;}
\newcommand{\Aachen}{III. Physikalisches Institut, RWTH Aachen University, 52056 Aachen, Germany}
\newcommand{\Alabama}{Department of Physics and Astronomy, University of Alabama, Tuscaloosa, Alabama 35487, USA}
\newcommand{\Argonne}{Argonne National Laboratory, Argonne, Illinois 60439, USA}
\newcommand{\APC}{APC, AstroParticule et Cosmologie, Universit\'{e} Paris Diderot, CNRS/IN2P3, CEA/IRFU, Observatoire de Paris, Sorbonne Paris Cit\'{e}, 75205 Paris Cedex 13, France}
\newcommand{\CBPF}{Centro Brasileiro de Pesquisas F\'{i}sicas, Rio de Janeiro, RJ, cep 22290-180, Brazil}
\newcommand{\Chicago}{The Enrico Fermi Institute, The University of Chicago, Chicago, IL 60637, USA}
\newcommand{\CIEMAT}{Centro de Investigaciones Energ\'{e}ticas, Medioambientales y Tecnol\'{o}gicas, CIEMAT, E-28040, Madrid, Spain}
\newcommand{\Columbia}{Columbia University; New York, NY 10027, USA}
\newcommand{\Davis}{University of California, Davis, CA-95616-8677, USA}
\newcommand{\Drexel}{Physics Department, Drexel University, Philadelphia, Pennsylvania 19104, USA}
\newcommand{\Hamburg}{Institut f\"{u}r Experimentalphysik, Universit\"{a}t Hamburg, 22761 Hamburg, Germany}
\newcommand{\Hiroshima}{Hiroshima Institute of Technology, Hiroshima, 731-5193, Japan}
\newcommand{\IIT}{Department of Physics, Illinois Institute of Technology, Chicago, Illinois 60616, USA}
\newcommand{\INR}{Institute of Nuclear Research of the Russian Aacademy of Science, Russia}
\newcommand{\CEA}{Commissariat \`{a} l'Energie Atomique et aux Energies Alternatives, Centre de Saclay, IRFU, 91191 Gif-sur-Yvette, France}
\newcommand{\Livermore}{Lawrence Livermore National Laboratory, Livermore, CA 94550, USA}
\newcommand{\Kansas}{Department of Physics, Kansas State University, Manhattan, Kansas 66506, USA}
\newcommand{\Kobe}{Department of Physics, Kobe University, Kobe, 657-8501, Japan}
\newcommand{\Kurchatov}{NRC Kurchatov Institute, 123182 Moscow, Russia}
\newcommand{\MIT}{Massachusetts Institute of Technology; Cambridge, MA 02139, USA}
\newcommand{\MaxPlanck}{Max-Planck-Institut f\"{u}r Kernphysik, 69029 Heidelberg, Germany}
\newcommand{\Niigata}{Department of Physics, Niigata University, Niigata, 950-2181, Japan}
\newcommand{\NotreDame}{University of Notre Dame, Notre Dame, IN 46556-5670, USA}
\newcommand{\IPHC}{IPHC, Universit\'{e} de Strasbourg, CNRS/IN2P3, F-67037 Strasbourg, France}
\newcommand{\SUBATECH}{SUBATECH, CNRS/IN2P3, Universit\'{e} de Nantes, Ecole des Mines de Nantes, F-44307 Nantes, France}
\newcommand{\Sussex}{Department of Physics and Astronomy, University of Sussex, Falmer, Brighton BN1 9QH, United Kingdom}
\newcommand{\Tennessee}{Department of Physics and Astronomy, University of Tennessee, Knoxville, Tennessee 37996, USA}
\newcommand{\TokyoInst}{Department of Physics, Tokyo Institute of Technology, Tokyo, 152-8551, Japan  }
\newcommand{\TokyoMet}{Department of Physics, Tokyo Metropolitan University, Tokyo, 192-0397, Japan}
\newcommand{\TohokuUni}{Research Center for Neutrino Science, Tohoku University, Sendai 980-8578, Japan}
\newcommand{\Muenchen}{Physik Department, Technische Universit\"{a}t M\"{u}nchen, 85747 Garching, Germany}
\newcommand{\TohokuGakuin}{Tohoku Gakuin University, Sendai, 981-3193, Japan}
\newcommand{\Tubingen}{Kepler Center for Astro and Particle Physics, Universit\"{a}t T\"{u}bingen, 72076, T\"{u}bingen, Germany}
\newcommand{\UFABC}{Universidade Federal do ABC, UFABC, Sao Paulo, Santo Andr\'{e}, SP, Brazil}
\newcommand{\UNICAMP}{Universidade Estadual de Campinas-UNICAMP, Campinas, SP, Brazil}
\newcommand{\Aviette}{Laboratoire Neutrino de Champagne Ardenne, domaine d'Aviette, 08600 Rancennes, France}
\newcommand{\deceased}{Deceased.}

\usepackage{graphicx}
\usepackage{dcolumn}
\usepackage{bm}

\begin{document}

\preprint{APS/123-QED}

\title{Indication for the disappearance of reactor \neb in the Double Chooz experiment}

\affiliation{\Aachen}
\affiliation{\Alabama}
\affiliation{\Argonne}
\affiliation{\APC}
\affiliation{\CBPF}
\affiliation{\Chicago}
\affiliation{\CIEMAT}
\affiliation{\Columbia}
\affiliation{\Davis}
\affiliation{\Drexel}
\affiliation{\Hamburg}
\affiliation{\Hiroshima}
\affiliation{\IIT}
\affiliation{\INR}
\affiliation{\CEA}
\affiliation{\Livermore}
\affiliation{\Kansas}
\affiliation{\Kobe}
\affiliation{\Kurchatov}
\affiliation{\MIT}
\affiliation{\MaxPlanck}
\affiliation{\Niigata}
\affiliation{\NotreDame}
\affiliation{\IPHC}
\affiliation{\SUBATECH}
\affiliation{\Sussex}
\affiliation{\Tennessee}
\affiliation{\TokyoInst}
\affiliation{\TokyoMet}
\affiliation{\TohokuUni}
\affiliation{\Muenchen}
\affiliation{\TohokuGakuin}
\affiliation{\Tubingen}
\affiliation{\UFABC}
\affiliation{\UNICAMP}

\author{Y.~Abe}
\affiliation{\TokyoInst}

\author{C.~Aberle}
\affiliation{\MaxPlanck}

\author{T.~Akiri}
\affiliation{\APC}
\affiliation{\CEA}

\author{J.C.~dos Anjos}
\affiliation{\CBPF}

\author{F.~Ardellier}
\affiliation{\CEA}

\author{A.F.~Barbosa}
\altaffiliation{\deceased}
\affiliation{\CBPF}

\author{A.~Baxter}
\affiliation{\Sussex}

\author{M.~Bergevin}
\affiliation{\Davis}

\author{A.~Bernstein}
\affiliation{\Livermore}

\author{T.J.C.~Bezerra}
\affiliation{\TohokuUni}

% ...


\collaboration{Double Chooz Collaboration}

\date{\today}

\begin{abstract}

The Double Chooz Experiment presents an indication of reactor electron
antineutrino disappearance consistent with neutrino oscillations.  An observed-to-predicted ratio of events of \mbox{0.944 $\pm$ 0.016~({\rm stat}) $\pm$ 0.040~({\rm syst})} was obtained in 101 days of running at
the Chooz Nuclear Power Plant in France, with two 4.25 GW$_{th}$ reactors.
The results were obtained from a single 10 m$^3$ fiducial volume detector
located 1050~m from the two reactor cores.  The reactor antineutrino flux 
 prediction used the Bugey4 flux measurement after correction for differences in core composition.    
The deficit can be interpreted as an indication of a non-zero value of the still unmeasured
neutrino mixing parameter \sang. Analyzing both the rate of the prompt positrons and their energy spectrum we find \mbox{\sang  = 0.086 $\pm$ 0.041~({\rm stat}) $\pm$0.030~({\rm syst})}, or, at 90\%~CL, \mbox{ 0.017 $<$ \sang $\ <$ 0.16}.

\end{abstract}


\maketitle




 We report first results of a search for a non-zero neutrino oscillation~\cite{PDG} mixing angle, \ang, based on reactor antineutrino disappearance. This is the last of the three neutrino oscillation mixing angles~\cite{PMNS1,PMNS2} for which only upper limits~\cite{Chooz,PaloVerde} are available. The size of \ang\  sets the required sensitivity of long-baseline oscillation experiments attempting to measure CP violation in the neutrino sector or the mass hierarchy. 

 In reactor experiments~\cite{Minakata,Min2} addressing the disappearance of \neb, \ang\  determines the survival probability of electron antineutrinos at the ``atmospheric" squared-mass difference, $\Delta m^2_{atm}$. This probability is given by:
\begin{equation}
P_{surv} \approx 1 - \sin^2 2\theta_{13} \sin^2 (1.267 \Delta m^2_{atm} L/E)\;,  \label{eq:osc}
\end{equation}
where $L$ is the distance from reactor to detector in meters and $E$ the energy of the antineutrino in MeV.  The full formula can be found in Ref.~\cite{PDG}. Eq.~\ref{eq:osc} provides a direct way to measure \ang\  since the only additional input is the well measured value of \mbox{$|\Delta m^2_{atm}| = (2.32^{+0.12}_{-0.08})\times 10^{-3}$ eV$^2$ \cite{MINOSDm2}}. Other running reactor experiments \cite{DayaBay,RENO} are using the same technique.  

Electron antineutrinos of \mbox{$<9$~MeV} are produced by reactors and detected through inverse beta decay (IBD):
\mbox{$\bar \nu_e + p \rightarrow e^+ + n$}. Detectors based on hydrocarbon liquid scintillators provide the free proton targets. The IBD signature is a coincidence of a prompt positron signal followed by a delayed neutron capture. The \neb energy, E$_{\bar \nu_e}$, is reconstructable from \Ep, the positron visible energy \mbox{(E$_{\bar \nu_e}$ $\cong$ \Ep + 0.78~MeV)}.

Recently, indications of non-zero \ang\  have been reported by two accelerator appearance experiments: T2K~\cite{T2kAbe} and MINOS~\cite{MINOSAdamson}. Global fits (see e.g.~\cite{Schwetz,Fogli}) indicate central values in the range  \mbox{$0.05 < \sin^2 2\theta_{13} < 0.10$}, accessible to the Double Chooz experiment~\cite{DC,Mention}.

We present here our first results with a detector located $\sim1050$~m from the two $4.25$~GW$_{th}$ thermal power reactors of the Chooz Nuclear Power Plant and under a $300$~MWE rock overburden. The analysis is based on $101$ days of data including $16$~days with one reactor off and one day with both reactors off.


The antineutrino flux of each reactor depends on its thermal power and, for the four main fissioning isotopes, $^{235}$U, $^{239}$Pu, $^{238}$U, $^{241}$Pu, their fraction of the total fuel content, their energy released per fission, and their fission and capture cross-sections. The fission rates and associated errors were evaluated using two predictive and complementary reactor simulation codes: MURE~\cite{MURE2,MURE-NEA} and DRAGON~\cite{DRAGON}.  
This allowed  a study of the sensitivity to the important reactor parameters ({\it e.g..} thermal power, boron concentration, temperatures and densities). The quality of these simulations was evaluated through benchmarks~\cite{TakahamaPaper}, and comparisons with Electricit\'{e} de France (EDF) assembly simulations. The maximum discrepancies observed were included in the fission rate systematic error.

MURE was used to develop a 3D simulation of the reactor cores. EDF provided the information required to simulate the fission rates including initial burnups of assemblies. 
To determine the inventories of each assembly composing the core at the startup of the data-taking cycle, assembly simulations were performed and the inventories at the given burnup computed. The energies per fission computed by Kopeikin~\cite{Kopeikin} and nuclear data evaluated from the JEFF3.1 database~\cite{JEFF} were used. 
The evolutions of the core simulations with time were performed using the thermal power and the boron concentration from the EDF database averaged over $48$~h time steps, yielding the relative contributions to fissions of the four main isotopes. 
 
    The associated antineutrino flux was computed using the improved spectra from~\cite{Huber}, converted from the ILL reference electron spectra~\cite{SchreckU5,SchreckU5Pu9,SchreckPu9Pu1}, and the updated {\it ab initio} calculation of the $^{238}\text{U}$ spectrum~\cite{Mueller2011}. 
The ILL spectra were measured after irradiating U or Pu for $\sim$ 1 day. Contributions from $\beta$-decays with lifetimes longer than these irradiation times were accounted for as prescribed in~\cite{Mueller2011}.

\begin{figure}[b]
  \includegraphics[scale=1.70]{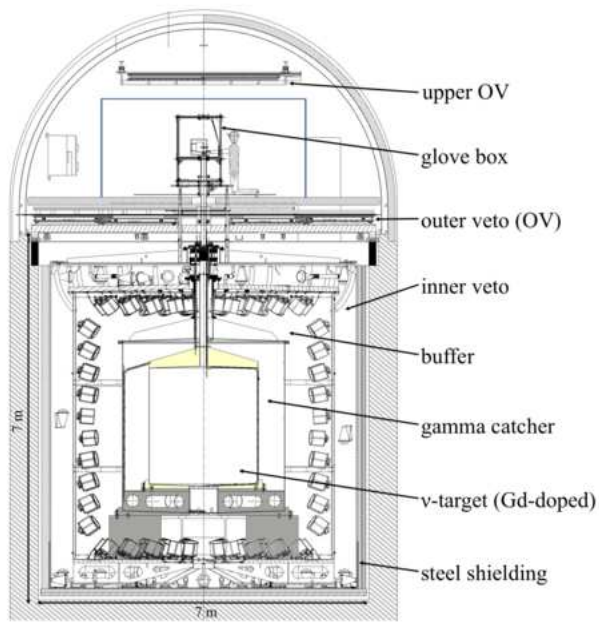}
  \caption{\label{fig:Det} A cross-sectional view of the Double Chooz detector system.}
\end{figure}

The Double Chooz detector system (Figure~\ref{fig:Det}) consists of a main detector, an outer veto, and calibration devices.
The main detector comprises four concentric cylindrical tanks filled with liquid scintillators or mineral oil. The innermost $8$~mm thick transparent (UV to visible) acrylic vessel houses the $10$~m$^{3}$ $\nu$-target liquid, a mixture of n-dodecane, PXE, PPO, bis-MSB and \mbox{1~g gadolinium/l} as a beta-diketonate complex. The scintillator choice emphasizes radiopurity and long term stability~\cite{sciprep}. The $\nu$-target volume is surrounded by the $\gamma$-catcher, a $55$~cm thick Gd-free liquid scintillator layer in a second $12$~mm thick acrylic vessel, used to detect \mbox{$\gamma$-rays} escaping from the $\nu$-target. The light yield of the $\gamma$-catcher was chosen to provide identical photoelectron (pe) yield across these two layers~\cite{scint}. Outside the $\gamma$-catcher is the buffer, a $105$~cm thick mineral oil layer.
It shields from radioactivity of photomultipliers (PMTs) and of the surrounding rock, and is one of the major improvements over the CHOOZ experiment~\cite{Chooz}.
$390$~10-inch PMTs~\cite{pmt,pmt2,pmt3} are installed on the stainless steel buffer tank inner wall to collect light from the inner volumes.
These three volumes and the PMTs constitute the inner detector (ID).
 
Outside the ID, and optically separated from it, is a $50$~cm thick ``inner veto'' liquid scintillator (IV).
It is equipped with $78$~8-inch PMTs and functions as a cosmic muon veto and as a shield to spallation neutrons produced outside the detector.
The detector is surrounded by $15$~cm of demagnetized steel to suppress external $\gamma$-rays.
The main detector is covered by an outer veto system (not used in this analysis).

The readout is triggered by custom energy sum electronics~\cite{trigger,trigger2,trigger3}.
The ID PMTs are separated into two groups of $195$~PMTs uniformly distributed throughout the volume and the PMT signals in each group are summed. The signals of the IV PMTs are also summed.
 If any of the three sums is above a set energy threshold,
the detector is read out with $500$~MHz flash-ADC electronics~\cite{fadc,fadc2} with customized firmware and a deadtime-free acquisition system.
Upon each trigger, a $256$~ns interval of the waveforms of both ID and IV signals is recorded.
The low trigger rate ($120$~Hz) allowed the ID readout threshold to be set at $350$~keV, well below the $1.02$~MeV minimum energy of an IBD positron,
greatly reducing the threshold systematics.

\begin{figure}[hb!]
  \includegraphics[scale=0.45]{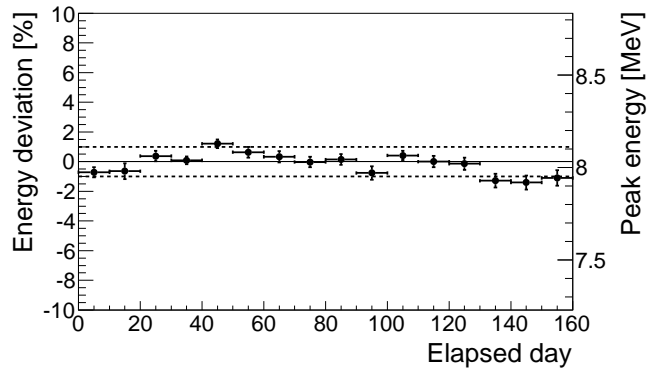}
  \caption{\label{fig:stability} The peak of the energy of neutron captures on Gd in IBD events(right scale) and its deviation from its average value(left scale) as a function of elapsed(calendar) day.}
\end{figure}

The experiment is calibrated by several methods.
A multi-wavelength  LED--fiber light injection system (LI) produces fast light pulses illuminating the PMTs from fixed positions.   
Radio-isotopes  $^{137}$Cs, $^{68}$Ge,
$^{60}$Co, and $^{252}$Cf were deployed in the target along the vertical symmetry axis 
and, in the gamma catcher, through a rigid loop traversing the interior
and passing along  boundaries with the target and the buffer. The detector was monitored using spallation neutron captures on H and
Gd, residual natural radioactivity, and daily LI runs. The stability of the peak energy of neutron captures on Gd in IBD candidates is shown in Figure~\ref{fig:stability}. The energy response was found to be stable within 1\% over time. 

The signature of IBD events is a delayed coincidence between a prompt positron energy
deposition, \Ep, and a delayed energy deposition, \Ed, due to the
neutron capture on H or Gd within \Delrel. The fiducial volume is constrained to the target vessel without position cuts by requiring a \neb event to have a capture on Gd, identified by its emission of $\sim$ 8 MeV in $\gamma$ rays.
The analysis compares the number and energy distribution of 
detected events to a prediction based on the
reactor data.

Energy measurements are based on the total charge, Q$_{tot}$,
collected by the PMTs and corrected for gain variations.    The energy is 
reconstructed scaling Q$_{tot}$ by a constant, adjusted so that the energy of the gamma emitted following neutron capture on H reconstructs to $2.22$~MeV
at the target center. This corresponds to $\sim 200$~pe/MeV. Our Monte Carlo (MC), based on GEANT4~\cite{Geant4}, is used to model the detector response and to calculate its acceptance. It uses 
parameters for quenching~\cite{quenAberle}, absorption, re--emission, refraction, etc. determined from laboratory  measurements of the detector liquids.  Comparisons between actual and simulated calibration
data were used to develop a parametric function to correct the simulation, and to assess the uncertainties in the  energy reconstruction. The function is a product of two factors. One, dependent on energy, ranges from 0.97 to 1.05 for \mbox{0.7-10.0 MeV}. The other, dependent on position, ranges from 0.94 to 1.00 over the target volume.

The following criteria are applied to select \neb candidates. Triggers within a $1000~\mu$s window following a
cosmic muon crossing the IV or the ID ($46$~s$^{-1}$) are rejected to
limit spallation neutron and cosmogenic backgrounds. 
This requirement is followed by five selections:
 1) a cut rejecting events caused by some sporadically
  glowing PMT bases, resulting in light localized to a few PMTs and spread out in time: Q$_{max}$/Q$_{tot}<$0.09 (0.06) for the prompt (delayed)
  energy and \mbox{rms(t$_{\rm start}$)$< 40$~ns}, where Q$_{max}$ is the maximum
  charge recorded by a single PMT and rms(t$_{\rm start}$) is the standard deviation of the times
  of the first pulse on each PMT; 2) \mbox{0.7 MeV$<$\Ep$<$12.2
  MeV};  3) \mbox{6.0 MeV $<$\Ed$ <$12.0 MeV}; 4)  \mbox{2 $\mu$s $<$ \Delrel $<$
  100 $\mu$s}, where the lower cut eliminates correlated noise and the upper cut is determined by the $\sim30~\mu$s capture time on Gd; 5) a multiplicity cut to reject correlated backgrounds defined as no additional valid trigger from $100~\mu$s preceding the prompt candidate to $400~\mu$s after it. 
 Applying selections (1-5) yields $4121$~candidates or $42.6\pm0.7$~events/day, uniformly distributed within the target, for an analysis live time of $96.8$~days. 


%
%

 Contributions from background events surviving these cuts have been estimated as follows. Uncorrelated coincidences result mainly from the random association of a prompt energy deposition due to radioactivity ($7.6$~s$^{-1}$) and a later candidate neutron capture \mbox({$\simeq$20/hour)}.
This background is measured by applying selection cuts
(1-5) but modifying selection (4) such that the $2-100~\mu$s time window is shifted by $1000~\mu$s relative to the prompt trigger. To improve the precision of this background measurement, $198$~such windows, each shifted from the previous one by $500~\mu$s, were used, leading to $0.33  \pm 0.03$~events per day. 

Fast neutrons induced by muons traversing 
the rock can interact in the target producing a recoil
proton and, later, be captured, simulating an
IBD event. We estimate this rate to be \mbox{$0.83\pm0.38$} events per day (including a contribution from stopping muons) 
by applying cuts (1-5), but modifying selection (2) such that \mbox{$12.2~$MeV $<$\Ep$<30$~MeV}, and then extrapolating to the signal region, assuming a flat
energy spectrum. We account for an uncertainty in this extrapolation, and for the contribution  of stopping muons, by including a shape error ranging up to \mbox{$\pm70$\%} of the flat extrapolation at lower energies. 

$^9$Li $\beta$-n emitters are produced preferentially by energetic muons.  They were studied by
searching for a triple delayed coincidence between a muon depositing $>600$~MeV in the detector and a \nebns-like pair of events, where the delay
between the muon and prompt event is dictated by the $178$~ms $^9$Li halflife, which precludes vetoing on all muons. Fitting the resulting time distribution with a flat component and an exponential with the $^9$Li lifetime results in an estimated rate of
 $2.3 \pm 1.2$~events/day. This rate is  assigned the energy spectrum of the $^9$Li decay branches. A shape uncertainty of up to 20\% is introduced to
 account for uncertainties in some decay branches.
$^8$He is not considered since it is less abundantly produced~\cite{KamIsotopes}. 
The total background rate, $3.46\pm1.26$~d$^{-1}$, is summarized in Table \ref{tab:background}. 

The overall background envelope is independently verified by analyzing
$22.5$~hours of both-reactors-off data  ($<0.3$~residual \neb events).
 Two \neb candidates, with prompt
energies of $4.8$~MeV and $9.8$~MeV, pass cuts (1-5). They were associated within 
$30$~cm and $220$~ms with
the closest energetic muon, 
 and are thus likely to be associated with $^9$Li.
\begin{table}[htb]
\caption{The breakdown of the estimated background rate. Additional shape uncertainties are described in the text.}
\label{tab:background}
\begin{tabular}{lcc}
\hline \hline
Background               & Rate/day & Syst. Uncertainty (\% of signal)\\
\hline
Accidental                  & $0.33  \pm 0.03$ & $<0$.1\\
\hline
Fast neutron              & $0.83\pm 0.38$   & 0.9\\
$^9$Li           & $2.3 \pm 1.2$   & 2.8\\
\hline \hline
\end{tabular}

\end{table}

 The following detector-related corrections and efficiencies as well as  their uncertainties were evaluated using the MC. The energy response  introduces a 1.7\% systematic uncertainty determined from fits to calibration data.
The number of free protons in the target scintillator,  
$6.747\times10^{29}$ based on its weight measurement, has an uncertainty of 0.3\%, originating from the knowledge of the scintillator hydrogen ratio.
A dedicated simulation including molecular bond effects~\cite{Tripoli} indicates
that the number of IBD events occurring in the gamma catcher with the
neutron captured in the target (spill in) exceeds the number of
events in the target with the neutron escaping to the gamma
catcher (spill out) by $1.4 \% \pm 0.4 \%$, 0.8\% lower than our standard MC prediction which was therefore reduced accordingly.
 Above the $700$~keV 
analysis threshold, the trigger efficiency is
$100.0^{+0}_{-0.4}$\%, assessed with a low threshold prescaled trigger. 
 Calibration data taken with the $^{252}$Cf source were used to check the MC
for any biases in the neutron selection criteria and estimate their contributions
to the systematic uncertainty.
The fraction of neutron captures on Gd is found to be
 ($86.0 \pm 0.5) \%$ near the center of the target, 2.0\% lower than the 
 simulation prediction which was reduced accordingly with a relative systematic uncertainty of 0.6\%.  The simulation reproduces the 96.5\% efficiency of the \Delrel cut with an uncertainty of 0.5\%
and the 94.5\% fraction of neutron captures on Gd accepted by the $6.0$~MeV cut with an uncertainty of 0.6\%.  
The MC normalization was adjusted for the muon veto
 ($-$4.5\%) and the multiplicity veto ($-$0.5\%) dead-times.

\begin{table}[htb!]
\caption{Contributions of the detector and reactor errors to the absolute normalization systematic uncertainty. }
\label{tab:systematics}
\begin{tabular}{lc|lc}
\hline \hline
\multicolumn{2}{c}{Detector} & \multicolumn{2}{c}{Reactor} \\
\hline \hline
Energy response &  $1.7$\%  & Bugey4 measurement & $1.4$\%  \\
\Ed $ \,$Containment &  $0.6$\% & Fuel Composition & $0.9$\%  \\
Gd Fraction         &  $0.6$\%  & Thermal Power & $0.5$\%   \\
\Delrel                 &  $0.5$\%  &Reference Spectra & $0.5$\% \\
Spill in/out           &  $0.4$\% & Energy per Fission & $0.2$\%  \\
Trigger Efficiency&  $0.4$\% & IBD Cross Section & $0.2$\%  \\
Target H              &  $0.3$\% & Baseline & $0.2$\%  \\
\hline 
Total                   &  $2.1$ \% & Total & $1.8$\%\\
\hline \hline
\end{tabular}

\end{table}

 The full covariance matrix of the emitted \neb spectra was computed as  prescribed in \cite{Mueller2011}. MURE provided the fractions of fissions per isotope $^{235}$U=48.8\%, $^{239}$Pu=35.9\%, $^{241}$Pu=6.7\%, and $^{238}$U=8.7\% and the fission rate covariance matrix. The resulting relative uncertainties on the above fission fractions are $\pm$3.3\%, $\pm$4\%, $\pm$11.0\% and $\pm$6.5\%, respectively. The error associated with the thermal power is \mbox{$\pm$0.46\%} at full power \cite{EDFTechnicalNote,AFNOR}, fully correlated between the two cores.

To avoid being affected by possible very short baseline \neb oscillations~\cite{RAA2011,Chooz,Giunti}, we adopt the reactor \neb spectrum of~\cite{Mueller2011, Huber}, but the global normalization is fixed by  the Bugey4 rate measurement~\cite{bugey} with its associated 1.4\% uncertainty. A relative correction of \mbox{($0.9\pm1.3$\%)} of the Bugey4 value accounts for the difference in core inventories. The IBD differential cross section is taken from~\cite{VogelBeacom99}, using \mbox{$881.5 \pm 1.5$~s~\cite{PDG}} as the neutron lifetime. The systematic uncertainties are summarized in Table~\ref{tab:systematics}.
 The expected no-oscillation number of \neb candidates is $4344\pm165$, including background.

\begin{figure}[hbt!]
\includegraphics[scale=0.44]{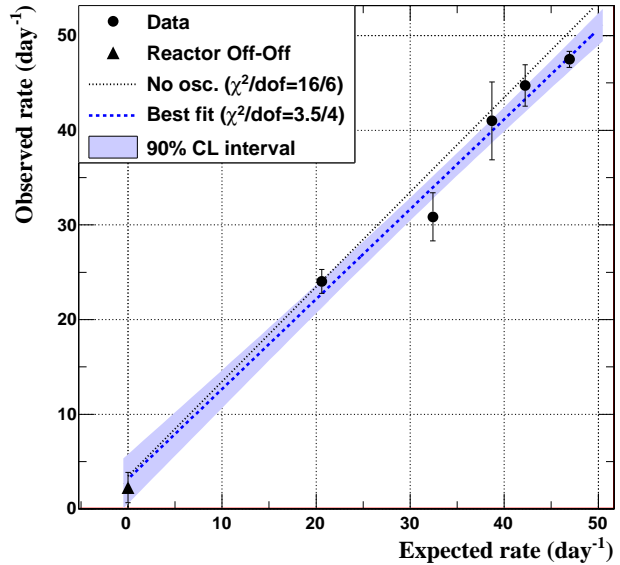}
\caption{\label{fig:nuyield}  Daily number of \neb candidates as a
  function of the expected number of  \neb.  The dashed line is a fit to the data, the band is the 90\% C.L. of this fit. The dotted line is the expectation in the no-oscillation
  scenario. The triangle indicates the measurement with both
  reactors off.}
\end{figure}

 The measured daily rate of IBD candidates as a
function of the no-oscillation expected rate for different reactor power conditions is shown in Figure~\ref{fig:nuyield}. The extrapolation to zero reactor power of the fit to the data (including the both-reactors-off) yields $3.2\pm1.3$~events per day, in excellent agreement with our background estimate and the both-reactors-off data.

Our measurement can be expressed as an observed IBD cross section per fission, \mbox{$\sigma^{DC}_{f}$}, a quantity which depends on the number of events observed, the number of target protons, the detector efficiency, the number of fissions occurring during our measurement and the distance to the reactors, yielding \mbox{$\sigma^{DC}_{f}=(5.383 \pm 0.210)\, 10^{-43}$ cm$^2$/fission}.  The Bugey4 measurement, corrected to match our fractions of isotopes quoted above, yields a cross section per fission of \mbox{$(5.703 \pm 0.108)\, 10^{-43}$ cm$^2$/fission}. The ratio of these two measurements is independent of any possible very short baseline oscilations. (Without Bugey4 normalization, the prediction, for our running conditions and using the reference spectra~\cite{Mueller2011,Huber}, is \mbox{$(6.209 \pm 0.170)\, 10^{-43}$ cm$^2$/fission}).

The ratio of observed to expected events is 
\mbox{$R_{DC}$ = 0.944 $\pm$ 0.016 ({\rm stat}) $\pm$ 0.040 ({\rm syst})}, 
corresponding to \mbox {\sang  = 0.104 $\pm$ 0.030 ({\rm stat}) $\pm$ 0.076 ({\rm syst})} for \mbox{$\Delta m^2_{13} = 2.4 \times 10^{-3}$ eV$^2$}.

The analysis is improved by comparing the positron spectrum in 
18 variably sized energy bins between \mbox {0.7 and 12.2~MeV} to the expected number of \neb events, again using \mbox{$\Delta m^2_{13} = 2.4 \times 10^{-3}$ eV$^2$}.
The analysis, performed with a standard $\chi^2$ estimator, uses four covariance matrices  to include uncertainties in
the antineutrino signal, detector response, signal and
background statistics, and background spectral shape. With very few
positrons expected above $8$~MeV, the region
$8-12.2$~MeV reduces the uncertainties in the correlated backgrounds with some additional contribution to the statistical uncertainty.  

\begin{figure}[hbt!]
\includegraphics[scale=0.42]{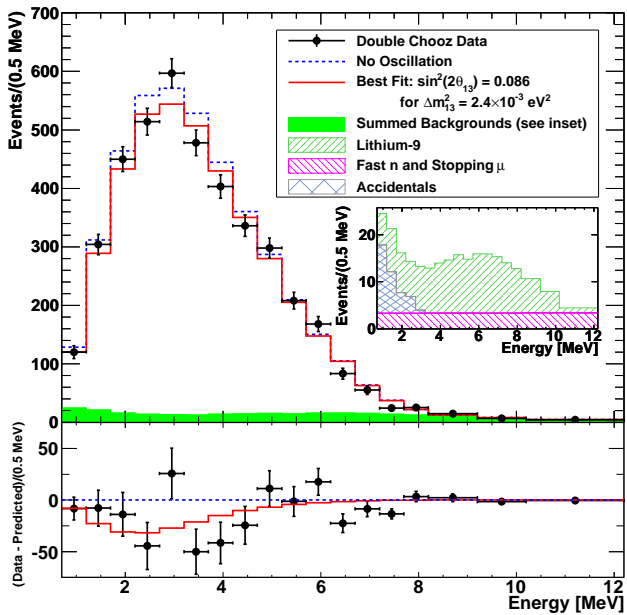}
\caption{\label{fig:e+spectra} Top: Expected prompt energy spectra,
  including backgrounds, for the
  no-oscillation case and for the best fit \sang, superimposed on the measured 
  spectrum. Inset: stacked histogram of backgrounds. Bottom: Difference between data and the no-oscillation spectrum (data points) and difference between the best fit and no-oscillation expectations (curve)}
\end{figure} 
The best fit results in \mbox{\sang  =  0.086 $\pm$ 0.041 ({\rm stat}) $\pm$ 0.030 ({\rm syst})} with a $\chi^2$/DOF of 23.7/17, whereas the \mbox{\sang  =  0.0} hypothesis results in a $\chi^2$/DOF of 26.6/18. Using a frequentist approach~\cite{FeldCous} we find an allowed region of \mbox{ 0.017 $<$ \sang $<$ 0.16} at 90\%~CL, and exclude the no oscillation hypothesis at the 94.6\%~C.L.

We determine our best estimate of the \neb and background
rates with a pulls-based approach~\cite{Pulls}, the results of which are shown in Table~\ref{tab:pullstable}. From the best fit we obtain 
a contribution from $^9$Li reduced by $\sim$19\%, and with an uncertainty decreased from 52\% to 26\%. The fast neutron value is decreased by 5\% with almost unchanged uncertainty.  
\begin{table}[htb]
\caption{ Summary of the effect of a pulls term approach on the fast neutron and $^9$Li backgrounds and on the energy scale.  Uncertainty values are in parentheses.}
\label{tab:pullstable}
\begin{tabular}{lccc}
\hline \hline
               & Fast n. Bkg(\%) &  $^9$Li (\%) &  EScale (value)\\
\hline
Rate only                 & 100 (46) & 100 (52) & 0.997 (0.007) \\
\hline
Rate + Shape              & 95.2 (38)  & 81.5 (25.5) & 0.998 (0.005)\\

\hline \hline
\end{tabular}

\end{table}

Figure~\ref{fig:e+spectra} shows the measured positron spectrum
superimposed on the expected spectra for the no-oscillation hypothesis and
for the best fit (including fitted backgrounds).

Combining our result with the T2K~\cite{T2kAbe} and MINOS~\cite{MINOSAdamson} measurements leads to \mbox{0.003 $<$ \sang $<$ 0.219}
at the 3$\sigma$ level.

In summary, Double Chooz has searched for \neb disappearance using a $10$~m$^3$ detector located $1050$~m from two reactors. 
A total of $4121$~events
were observed where \mbox{4344 $\pm$ 165} were expected for no-oscillation, with a signal to background
ratio of $\approx$11:1. In the
context of neutrino oscillations, this deficit leads to \mbox{\sang = 0.086  $\pm$ 0.041 ({\rm stat}) $\pm$ 0.030 ({\rm syst})}, based on
an analysis using rate and energy spectrum information.
The no-oscillation hypothesis is ruled out at the 94.6\%~C.L.
Double Chooz continues to run, to reduce statistical and background systematic uncertainties. A near detector
 will soon lead to reduced reactor and detector
systematic uncertainties and to an estimated 1$\sigma$ precision on \sangsp of \mbox{$\sim$ 0.02}.
%

 We thank all the technical and administrative people who helped build the experiment and the CCIN2P3 computer center for their help and availability. We thank, for their participation, the French electricity company EDF, the European fund FEDER, the R\'{e}gion de Champagne Ardenne, the D\'{e}partement des Ardennes and the Communaut\'{e} des Communes Rives de Meuse. We acknowledge the support of CEA and CNRS/IN2P3 in France, MEXT and JSPS of Japan, the Department of Energy and the National Science Foundation of the United States, the Ministerio de Ciencia e Innovaci\'{o}n (MICINN) of Spain, the Max Planck Gesellschaft and the Deutsche Forschungsgemeinschaft DFG (SBH WI 2152), the Transregional Collaborative Research Center TR27, the Excellence Cluster "Origin and Structure of the Universe" and the Maier-Leibnitz-Laboratorium Garching, the Russian Academy of Science, the Kurchatov Institute and RFBR (the Russian Foundation for Basic Research), the Brazilian Ministry of Science, Technology and Innovation (MCTI), the Financiadora de Estudos e Projetos (FINEP), the Conselho Nacional de Desenvolvimento Cient\'{i}fico e Tecnol\'{o}gico (CNPq), the S\~{a}o Paulo Research Foundation (FAPESP) and the Brazilian Network for High Energy Physics (RENAFAE) in Brazil.



\bibliographystyle{apsrev}

\bibliography{./PubMar13PRL.bib}

\end{document}